\documentclass[
    ,final            
  ]
  {aipproc}

\layoutstyle{6x9}

\begin{document}

\title{After Acoustic Peaks: What's Next in CMB?}

\author{Asantha Cooray\footnote{Sherman Fairchild Senior Research Fellow}}{
  address={Theoretical Astrophysics Including Relativity Group, California Institute of Technology, Pasadena, California 91125. E-mail: asante@caltech.edu}
}



\begin{abstract}
The advent of high signal-to-noise cosmic microwave background (CMB) anisotropy experiments 
has allowed detailed studies on the power spectrum of 
temperature fluctuations. The existence of acoustic oscillations in the anisotropy power 
spectrum is now established with the detection of 
the first two, and possibly the third, peaks. Beyond the acoustic peak structure, 
we consider cosmological and astrophysical information that can be 
extracted by pushing anisotropy observations to fine angular scales with 
higher resolution instruments. At small scales, a variety of contributions 
allow the use of CMB photons as a probe of the large scale structure:
we outline possible studies related to understanding detailed physical properties such 
as the distribution 
of dark matter, baryons and pressure, and ways to measure the peculiar, transverse 
and rotational velocities
of virialized halos such as galaxy clusters. Beyond the temperature, 
we consider several useful aspects of the CMB polarization
and comment on an ultimate goal for future CMB experiments involving the direct 
detection of inflationary gravity-waves through their distinct signature in the curl-type polarization.
\end{abstract}

\maketitle


\section{CMB: At Present}

The cosmic microwave background (CMB) is now a well known probe of the early universe. 
The temperature fluctuations in the CMB, especially the so-called acoustic peaks in the angular power spectrum of CMB anisotropies, 
capture the physics of primordial photon-baryon fluid undergoing oscillations in the potential 
wells of dark matter (Hu et al. 1997). The associated physics --- involving the evolution of a single
photon-baryon fluid under Compton scattering and gravity --- are both simple and linear, and
many aspects of it have been discussed in the literature since the early 1970s (Peebles \& Yu 1970; Sunyave \& Zel'dovich 1970).
The gravitational redshift contribution at large angular scales (e.g., Sachs \& Wolfe 1968) and the photon-diffusion damping
at small angular scales (e.g., Silk 1968) complete this description.

\begin{figure}[t]
\includegraphics[height=25pc,angle=-90]{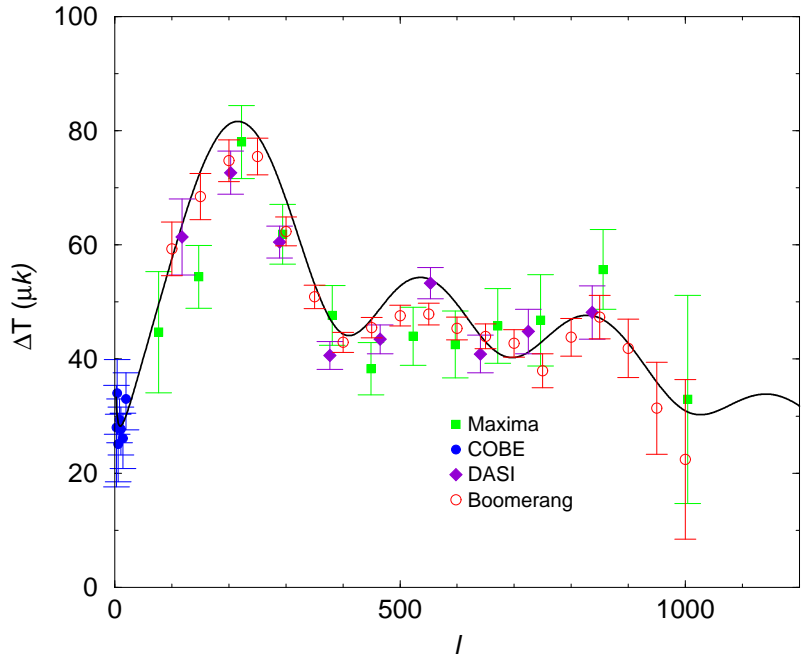}
\caption{Temperature fluctuations in the cosmic microwave background as seen by COBE (filled circles; Tegmark 1996),
Boomerang (open circles; Netterfield et al. 2001), MAXIMA (squares; Lee et al. 2001) and DASI (diamonds; Halverson et al. 2001).
The solid line shows the theoretical expectation for a cosmology with best-fit parameters for the Boomerang data following Netterfield et al. 2001}
\label{fig:cl}
\end{figure}

By now, there are at least five independent detections of the
first and second, and possibly the third, acoustic peak 
in the anisotropy power spectrum (Miller et al. 1999; de Bernardis et al. 2000; Hanay et al. 2000; Halverson et al. 2001).  We summarize 
these results in figure~\ref{fig:cl}.
Given the variety of experiments that are either collecting data or reducing  data that were recently collected,
 more detections that extend to higher peaks are soon expected. The NASA's MAP mission\footnote{http://map.gsfc.nasa.gov}
is expected to provide a significant detection of the acoustic peak structure out to a multipole of $\sim$ 1000 and, in the long term,
ESA's Planck surveyor\footnote{http://astro.estec.esa.nl/Planck/}, 
will extend this to a multipole of $\sim$ 2000 with better frequency coverage and polarization sensitivity.

A well known geometrical  test related to the CMB temperature 
anisotropy power spectrum involves the location of the first acoustic peak in the multipolar space.
The location depicts the projected sound horizon at the recombination and with increasing 
curvature, the location moves to smaller angular scales or higher $l$-values (Kamionkowski et al. 1994). 
The current data, with the first peak at a multipole of $\sim$ 200, strongly suggest a flat-universe (e.g., Jaffe et al. 2001). 
Additionally, alternative observations, such as the
baryon fraction in galaxy clusters and details related to the large scale structure clustering, 
have provided strong evidence for a low matter content in the universe (about 30\% to 40\% of the critical density; see 
Turner 2001 for a summary). Note that this matter content is significantly above the baryon contribution; the latter amounts 
to be about 5\% of the critical density based on calculations related to the big-bang
nucleosynthesis and the observed abundance of light elements.  The difference 
is the dark matter problem in astrophysics today.

Though one can effectively {\it invent} another form of an energy density in the
universe which does not clump and does not behave as matter to preserve the geometrical flatness, there is a consistent picture
from low redshift observations: the luminosity-distance diagram for type Ia supernovae 
out to a redshift of $\sim$ 1,  and higher, indicates that the universe is accelerating (e.g., Perlmutter et al. 1999).
 This result implies the presence of an additional energy component in the form of the so-called cosmological constant or, 
more generally, in the form of a dark energy or a quintessence. 
This dark energy has a large negative pressure responsible for the anti-gravity behavior. 
A complete particle physics description of the dark energy
still remains to be worked out. The physical nature of the dark energy presents the second important 
problem in astrophysics today.

In addition to these simple inductions on cosmology and detailed numerical estimates on parameters that define the
modern cosmological models, one can also note few basic things. For example, the acoustic oscillations 
in the CMB anisotropy power spectrum require phases of all Fourier modes to be the same at the beginning. The inflation is expected to
provide necessary adiabatic initial conditions while alternative models for structure formation, such as cosmic defects,
are now strongly ruled out by the data as they predict, at most,
 just one broad peak in the anisotropy power spectrum due to no coherence.
The shape of the first acoustic peak is also inconsistent with such alternative models and is more consistent with 
cold dark matter cosmologies favored  by many today. 

\section{CMB: As a Probe of the Local Universe}

\begin{figure}[t]
\includegraphics[height=20pc,angle=-90]{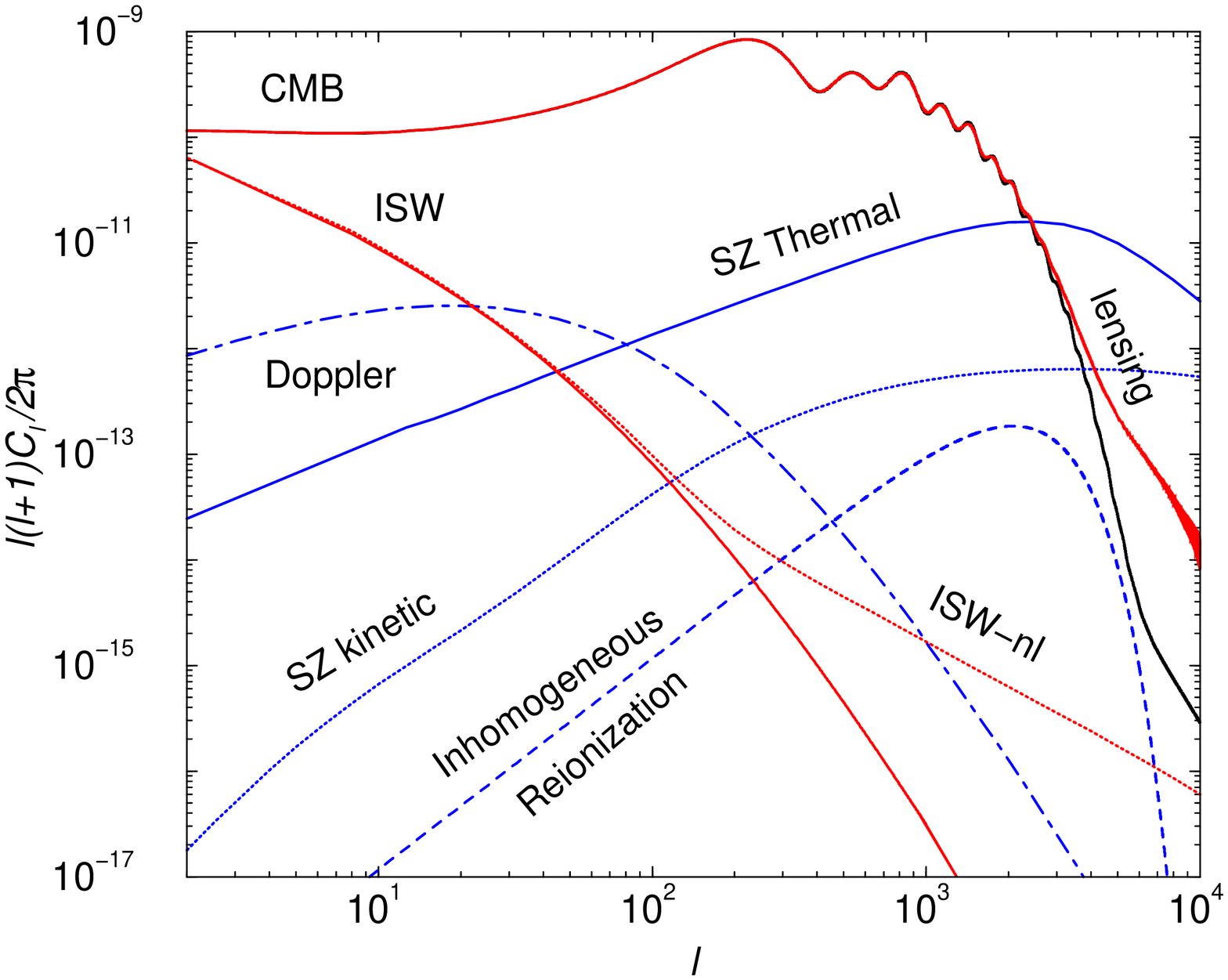} \includegraphics[height=20pc,angle=-90]{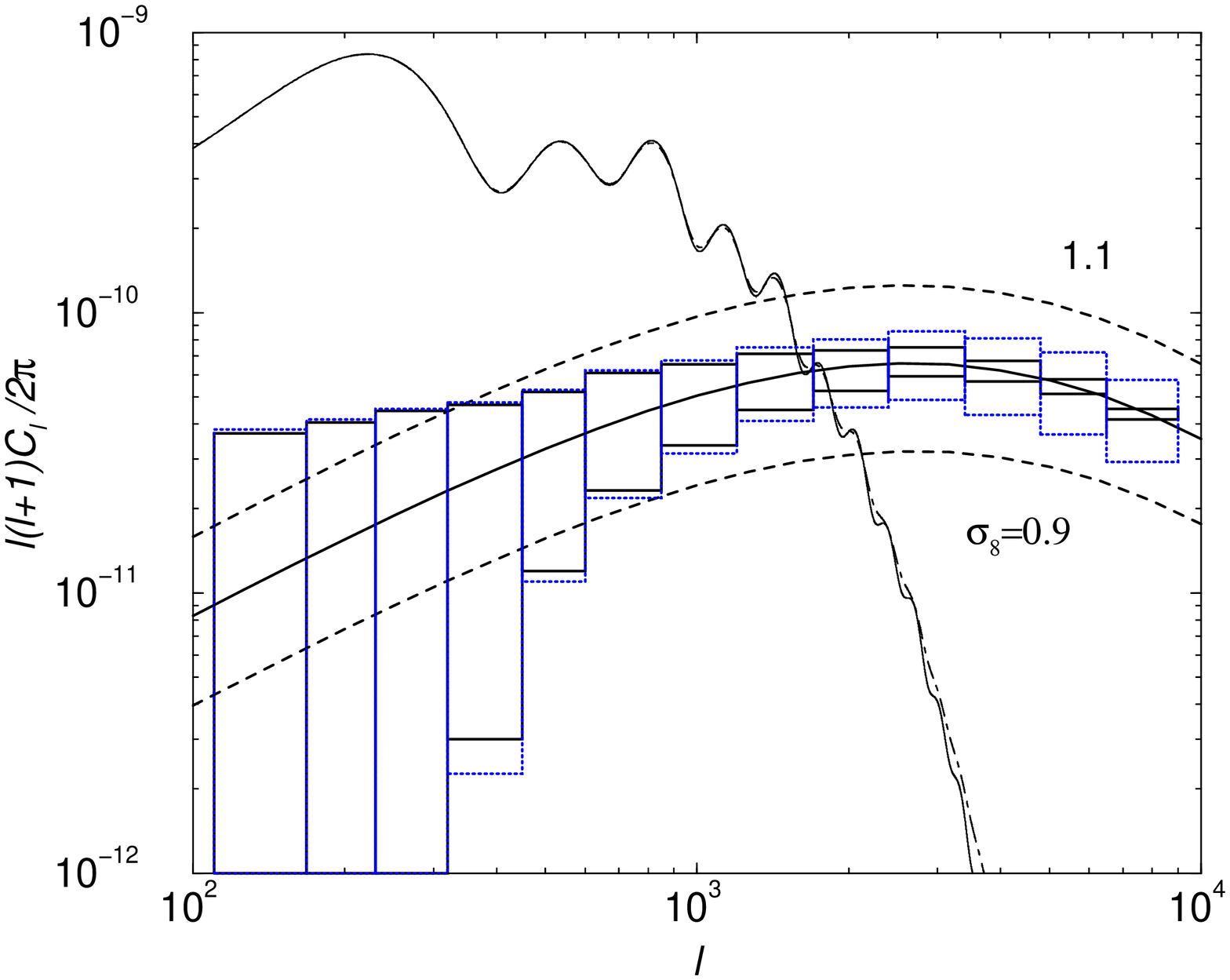}
\caption{{\it Left}: Power spectrum for the temperature
anisotropies in the fiducial $\Lambda$CDM model
with $\tau=0.1$. The curves show the local universe contributions to CMB due to
gravity (ISW and lensing) and scattering (Doppler, SZ effects, patchy reionization).
{\it Right:} The SZ thermal contribution in  focus: the three curves shows the variation in the SZ effect due to a change in
normalization about $\sigma_8=1.0$. Due to the highly non-linear behavior the SZ thermal contribution is strongly dependent on the normalization.
The two sets of error bars again show the highly non-linear and Poisson behavior of the SZ contributing clusters: the small
errors bars with sold likes are those under the Gaussian assumption only, while the dotted errors depict the total error with non-Gaussianities
included. The non-Gaussianities generally increase the power spectrum errors by factor of few at arcminute scales. We have assumed
a survey of 1 deg.$^2$ and no instrumental noise contribution to the power spectrum.}
\label{fig:sec}
\end{figure}

In transit to us, CMB photons also encounter the large scale structure that defines the local universe; thus, several 
aspects of photon properties, such as the frequency or the direction of propagation, are affected.
In the reionized epoch, variations are also 
imprinted when photons are  scattered via electrons, moving with respect to the CMB.
Though these secondary contributions to the CMB temperature are also generated by the same two processes 
that led to primary anisotropies, gravity and Compton-scattering (see, figure~\ref{fig:sec}),
there is one significant difference between the simple
linear description involving acoustic oscillations and the late-time modifications. In the latter case, one deals with the
complex and highly non-linear large scale structure. This involves detailed astrophysics at late times, such as the
evolution of gas or pressure, or the formation of structures that define the local universe.
Though these secondary effects are in some cases insignificant compared to primary fluctuations, they
leave certain imprints in the anisotropy structure and induce higher order correlations. 
These signatures can then be used to extract basic properties of the large scale structure that led to these 
additional changes to CMB temperature. 
Now, we will summarize what these signatures are, and, how they can be used to study the local universe with CMB photons.
 
{\bf Integrated Sachs-Wolfe Effect:}  The differential redshift effect from photons climbing in and
out of a time-varying gravitational potential along the line of
sight is called the integrated Sachs-Wolfe (ISW; Sachs \& Wolfe 1967) effect.
The ISW effect is important for low matter density universes
$\Omega_m < 1$, where the gravitational potentials decay at low redshift,
and contributes anisotropies on and above the scale of the horizon at the
time of decay. The non-linear contribution to the ISW effect comes from the large scale structure
momentum density field. In Cooray (2002), we presented a model for the non-linear contribution based on a simple
halo based description of the dark matter distribution (see, Cooray \& Sheth 2002 for a review).
A useful aspect of this non-linear effect is that it results from the transverse velocity of halos
across the line of sight and leads to a dipolar temperature change aligned with the direction of motion. 
Though this dipolar temperature distribution towards known galaxy clusters allow this effect to be easily distinguished,
a similar temperature pattern is also produced by the weak lensing effect involving the gradient of the cluster potential.
Since the velocity direction on the sky need not be that of the large scale CMB gradient, which is lensed, one can
effectively use a filtering scheme to separate out and extract the transverse velocity contribution from the lensing signature.

This non-linear transverse velocity contribution, however, 
produces fluctuations which are of order few $\mu$K when transverse velocities are of order 300 km sec$^{-1}$. 
One can attempt to reach the required subarcminute resolution required for such an observations by outfitting the
focal plane of the new 100m Green Bank Telescope with a bolometer array. Assuming a sensitivity of 200 $\mu$K$\sqrt{\rm sec}$
for each element, a 1000 element bolometer array can detect the effect, say towards the nearby Coma galaxy cluster,
in $\sim$ 100 hours; this should be compared to the similar, or more, integration times that are currently taking to 
produce images of the thermal Sunyaev-Zel'dovich effect in galaxy clusters at the BIMA and OVRO arrays 
by Carlstrom et al. (1996) with 1-sigma noise contributions of
few tens of $\mu$K. The ability to measure transverse velocities, which is generally not possible with
other astrophysical methods such as spectroscopy, challenges improvements in the experimental front including
novel observational techniques.

\begin{figure}[t]
\includegraphics[height=20pc,angle=-90]{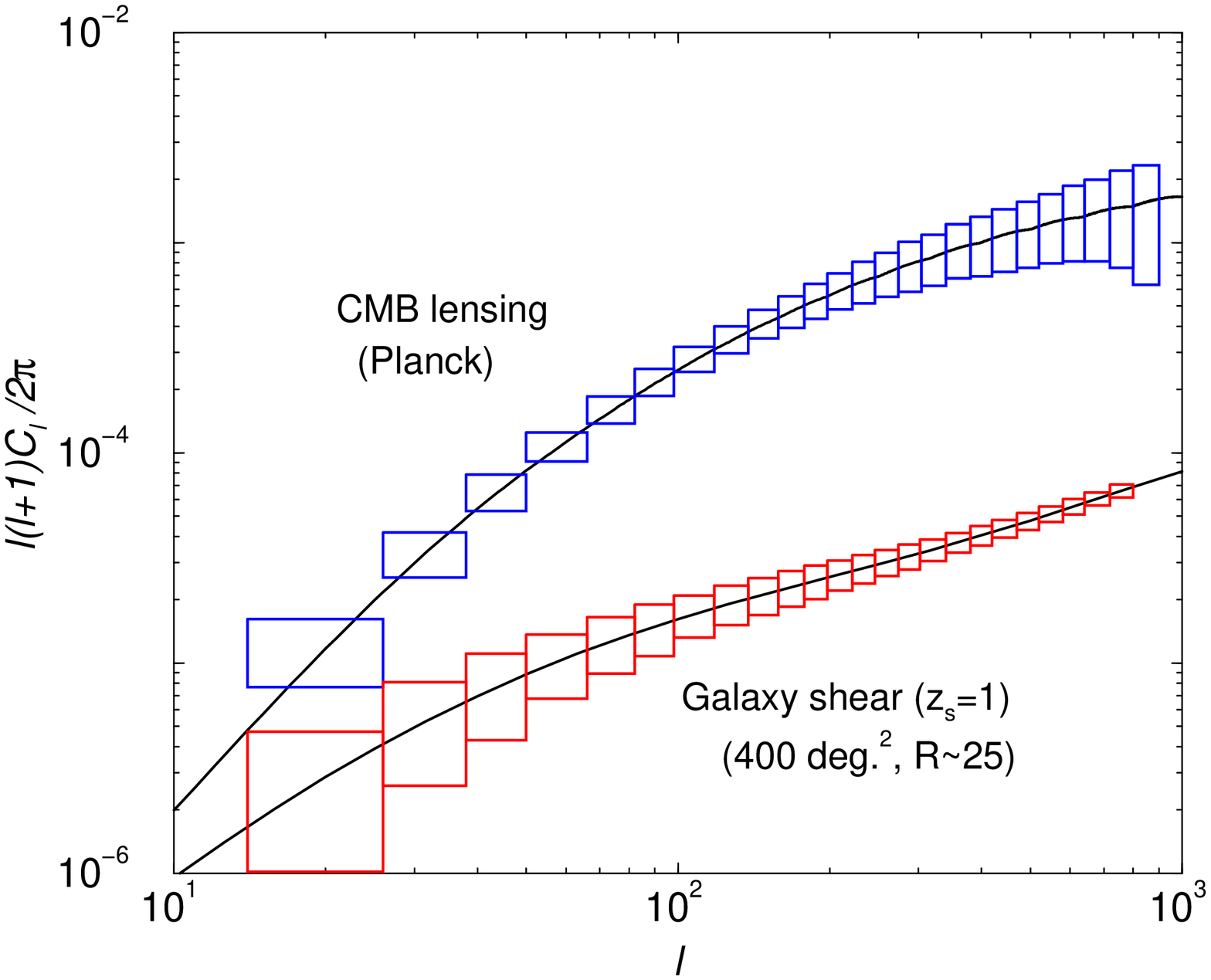}
\includegraphics[height=20pc,angle=-90]{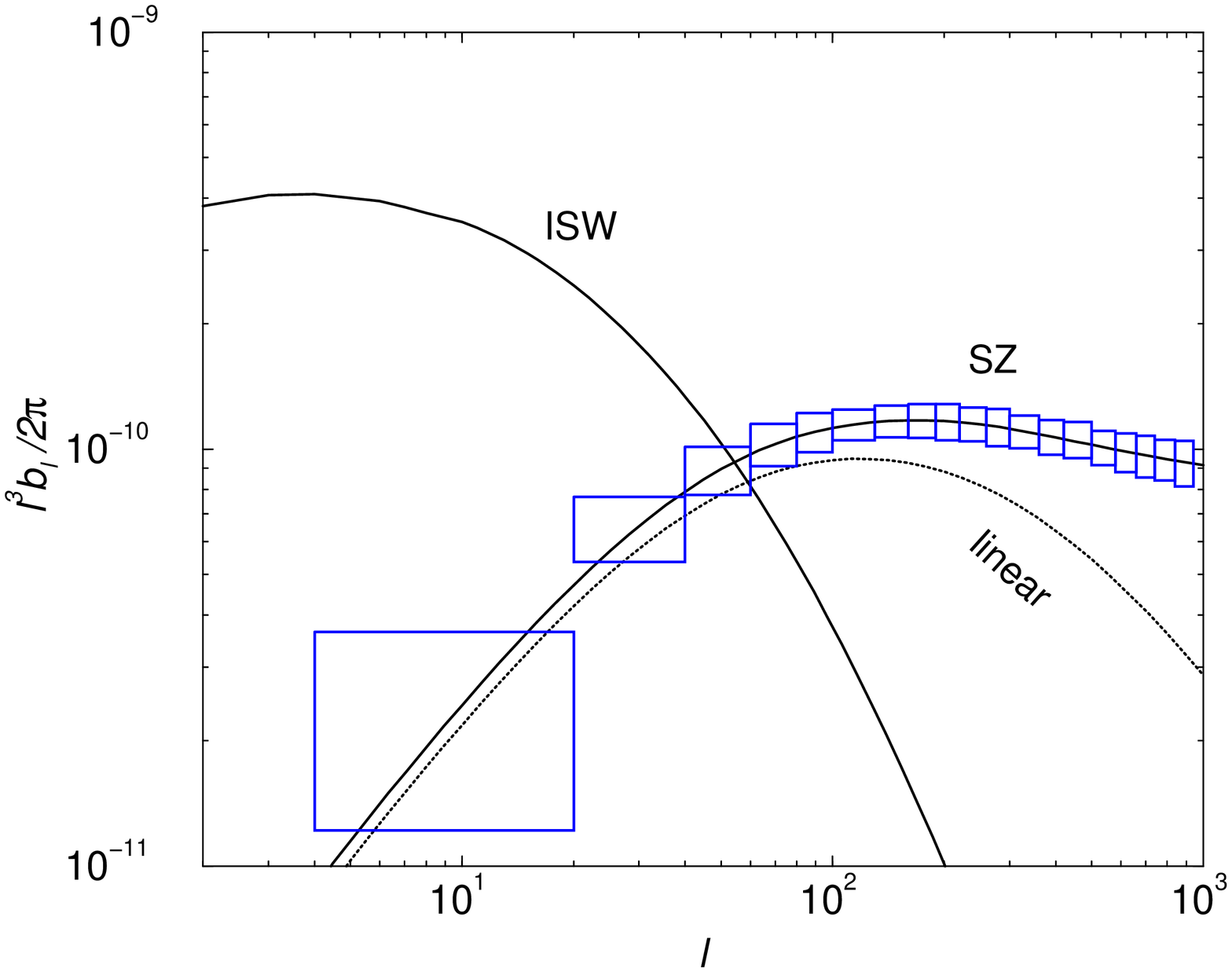}
\caption{{\it Left}: Weak lensing as a CMB experiment. We show the errors for reconstruction of convergence, or projected mass
density, to the last scattering surface with Planck temperature data. This is compared to the convergence measurements
expected from observations of galaxy ellipticities and probe to redshifts of $\sim$ 1 to 2.
{\it Right}: The construction of the dark matter-pressure correlation using the cross-correlation between lensing effect on
CMB and the SZ effect. The statistic used here is the CMB$^2$-SZ power spectrum.}
\label{fig:convergence}
\end{figure}

{\bf Weak Gravitational Lensing:} Gravitational lensing of the photons by the intervening large-scale structure both redistributes power in multipole space 
and enhances it due to power
in the density perturbations.  The most effective structures
for lensing lie half way between the surface of recombination and
the observer in comoving angular diameter distance.  In the
fiducial $\Lambda$CDM cosmology, this is at $z \sim 3.3$,
but the growth of structure skews this to
somewhat lower redshifts.  In general, the efficiency of lensing is
described by a broad bell shaped function between the source and the
observer, and thus, correlates well with a large number of tracers
of the large scale structure  from low to high redshifts.
 
Since the lensing effect involves the angular gradient of CMB photons and leaves the surface brightness unaffected, its  signatures are at 
the second order in temperature. Effectively, lensing smooths the acoustic peak structure at large angular scales and
moves photons to small scales. When the CMB gradient at small angular scales are lensed by foreground structures 
such as galaxy clusters, new anisotropies are generated at arcminute scales. For favorable cosmologies, the mean gradient is of order
$\sim$ 15 $\mu$K arcmin$^{-1}$ and with deflection angles or order 0.5 arcmin or so from massive clusters, the lensing effect results
in temperature fluctuations of order $\sim$ 5 to 10 $\mu$K. One can effectively extract this lensing contribution, and the
integrated dark matter density field responsible for the lensing effect, via 
 quadratic statistics in the temperature and the polarization.
This is best achieved with the divergence of the temperature weighted temperature gradient statistic of Hu (2001)
and we show errors for a construction of convergence in figure~\ref{fig:convergence}(a).
Since lensing  has a distinct non-linear mode coupling behavior,
it can produce a non-Gaussian signature in CMB data. The idea behind here is that the potentials which lensed CMB also contribute to 
first order temperature fluctuations from effects such as
the ISW effect.  These correlations are discussed in detail in Goldberg \& Spergel (1999) and Cooray \& Hu (2000).

\begin{figure}[t]
\includegraphics[height=15pc]{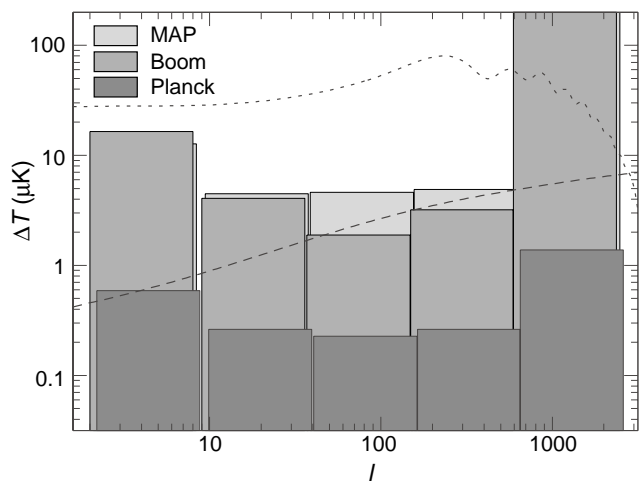}
\caption{Detection thresholds for the SZ effect.   Error boxes represent the
1-$\sigma$ rms residual noise in multipole bands and can be interpreted
as the detection threshold for MAP, Planck and Boomerang experiments.  Also shown (dotted) is the level of the primary
anisotropies that have been subtracted with the technique and the signal
(dashed) expected in simplified models for the large scale structure SZ effect.}
\label{fig:sz}
\end{figure}

{\bf Sunyaev-Zel'dovich Thermal Effect:} At small angular scales the best  known secondary contribution 
is the thermal Sunyaev-Zel'dovich
(SZ; Sunyaev \& Zel'dovich 1980) effect.
The SZ effect arises from the  inverse-Compton scattering of CMB photons by hot electrons
along the line of sight. This effect has now been directly imaged
toward massive galaxy clusters, where temperature of the scattering medium  can reach as high as 10 keV producing temperature changes in the CMB of order 
1 mK. Note that the effect is proportional to the temperature weighted electron density, or
pressure in the large scale structure. Using a distribution of halos with gas in hydrostatic equilibrium, 
we can calculate the pressure clustering and project it along the line of sight to obtain the resulting angular power spectrum of temperature 
fluctuations (Cooray 2000; 2001a). 

The general result from these halo based calculations, and confirmed
in numerical simulations, is that the large scale structure pressure is heavily dominated by  massive halos such as
galaxy clusters.  Thus, the SZ power strongly depends on the normalization of the power spectrum, such as the $\sigma_8$.
The recent detection of excess power at small scales by the Cosmic Background Imager (CBI) is consistent with a high $\sigma_8$ ($\sim$ 1),
but there is an additional effect involving the high non-Gaussianity. In figure~\ref{fig:sec}(b), we
show error bars on the  
SZ power spectrum from observations in a one square degree field. The smaller error bars assume Gaussian statistics, however, in
reality, the non-Gaussian aspect of the SZ contribution increases errors by a factor of few at arcminute scales. This is
equivalent to the statement that the SZ contribution varies significantly
 depending on whether one sees a massive cluster in the field of
observations or not. Since massive clusters are rare, observations which are limited to small areas on the sky are
biased. Any future interpretation of the SZ contribution should keep this increase in error, and any bias, in mind.
To effectively obtain a fair sample of the universe, one
has to survey 100 sqr. degrees or more instead of current surveys limited at most to few tens of square degrees.
 
In addition to these non-Gaussian aspects, the SZ thermal contribution should also be separated from the dominant primary fluctuations.
Note that the SZ effect also bears a spectral signature that differs from other effects.
The upscattering of photons moves them from low to high frequencies, with no effect at
a frequency of $\sim$ 217 GHz. This leads to decrements at low frequencies and increments at high frequencies. An experiment such as 
the Planck surveyor with sensitivity
beyond the peak of the spectrum can separate out the SZ contribution
based on the spectral signature (Cooray et al. 2000). In figure~\ref{fig:sz}, we summarize our results.
This frequency separation is 
important since statistics related to the SZ effect can be studied separately uncontaminated by primary anisotropies and confusing foregrounds. 
For example, in Cooray (2001b), we introduced the CMB$^2$-SZ power spectrum as a probe of the 
lensing-SZ correlation which involves the CMB and frequency cleaned SZ maps.
This statistic can be used to directly estimate how pressure correlates with dark matter;
in figure~\ref{fig:convergence}(b), we show expected errors for the Planck mission.
 
{\bf Linear Doppler Effect:}
The bulk flow of the electrons that scatter CMB photons leads to
a Doppler effect.  Its effect on the power spectrum  peaks around the
horizon at the scattering event projected on the sky
today (see figure~\ref{fig:sec}).
On scales smaller than the horizon at scattering, the contributions are
significantly canceled as photons scatter against the crests and troughs of
the perturbation.  As a result, the Doppler effect is moderately sensitive
to how rapidly the universe reionizes since contributions from a sharp
surface of reionization do not cancel.
 
{\bf Kinetic Sunyaev-Zel'dovich Effect:} 
The Ostriker-Vishniac/kinetic SZ (kSZ) effect arises from the second-order modulation of the Doppler effect by density fluctuations 
(Ostriker \& Vishniac 1986).   Due to the density weighting,
kSZ effect peaks at small scales  and avoids cancellations
associated with the linear effect; the linear effect can also be modulated via
the fraction of ionized electrons (Aghanim et al. 1996).
Due to the density modulation, the kSZ effect has a
distinct non-Gaussian behavior (Cooray 2001a). The bulk velocities projected along the line of sight 
can be extracted with higher order correlations such as the SZ$^2$-kSZ$^2$ statistic of Cooray (2001a).
For a given cluster, the temperature fluctuation
is proportional to the line of sight peculiar velocity and when the 
SZ thermal contribution is removed with frequency information, the kinetic SZ effect should dominate.
With multifrequency observations of galaxy clusters at high resolution, it should, in principle, be possible to
extract the bulk motions of clusters from the relative contributions of the thermal and kinetic SZ effects.
In addition to the peculiar motion, any coherent rotation of the cluster will produce 
a dipole-like temperature distribution toward clusters, 
especially when the rotational axes is aligned perpendicular to the line of sight (Cooray \& Chen 2001). 
One can again use filtering techniques to extract this dipolar signature due to the
gas distribution spanning a small extent when compared to the gradient of the potential that leads to effects such as lensing.
The dipolar pattern involved here allows an extremely useful 
probe of the cluster rotation and the angular momentum of gas distribution within galaxy clusters.

\begin{figure}[t]
\includegraphics[height=25pc,angle=-90]{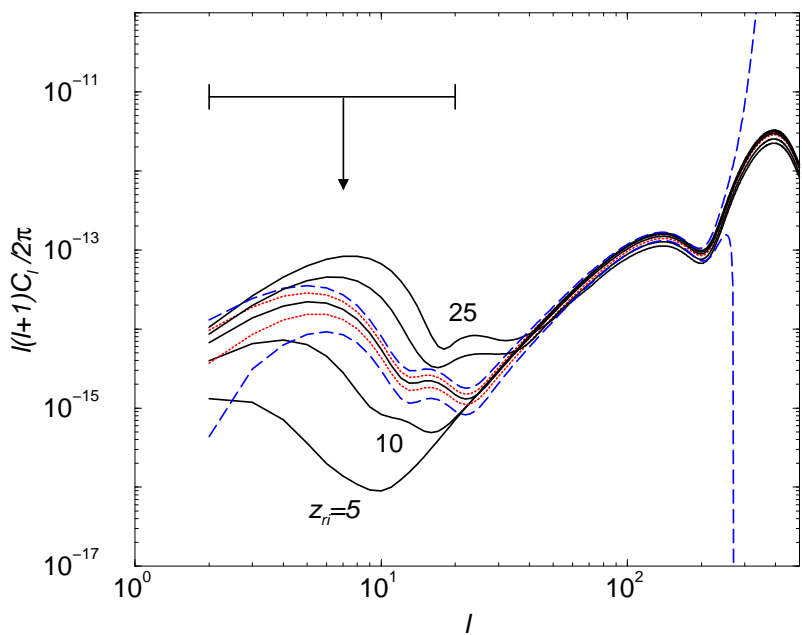}
\caption{The anisotropies in the polarization. We show the power spectrum of fluctuations in the E-mode for reionization from 5 to 25 at steps of 5.
The dotted lines are the expected errors for an all sky experiment, while the dashed lines are the expected errors for an
experiment with 100 $\mu$K $\sqrt{\rm sec}$ sensitivity and a  beam of 2.5 degrees and 25\% sky coverage. Such a dedicated experiment is possible in
the near future from sites such as South Pole. For comparison, we also show a current upper limit on the polarization signal from Keating et al. (2001).}
\label{fig:ee}
\end{figure}

{\bf Reionization via Polarization:} In addition to temperature anisotropies at smaller scales, there is also a push to observer polarization at medium to larger scales.
Though polarization contribution at the recombination peaks at $l \sim 1000$, the large scale polarization signal again allows a useful  probe of
the local universe and associated astrophysics. When the universe reionizes at low redshifts, currently believed to be
some where between redshifts of 6 and 25, the temperature quadrupole rescatters at the new reionization scattering surface 
and produces a new contribution
to the polarization.  This effect leads to a new peak in the polarization anisotropy power spectra
at very large angular scales and this peak scales as the square-root
of the reionization redshift (Zaldarriaga 1997). In figure~\ref{fig:ee}, we show the large angular scale polarization in the gradient, 
or the E,-mode. Since the signal is at large scales, one can easily construct a low resolution dedicated experiment 
for detection of the peak due to
rescattering. We show expected errors for an experiment with 25\% sky coverage, 100 $\mu$K$\sqrt{\rm sec}$ sensitivity and a 
beam of 2.5 degrees.  

\section{CMB: As a Probe of the Inflation}

\begin{figure}[t]
\includegraphics[height=20pc]{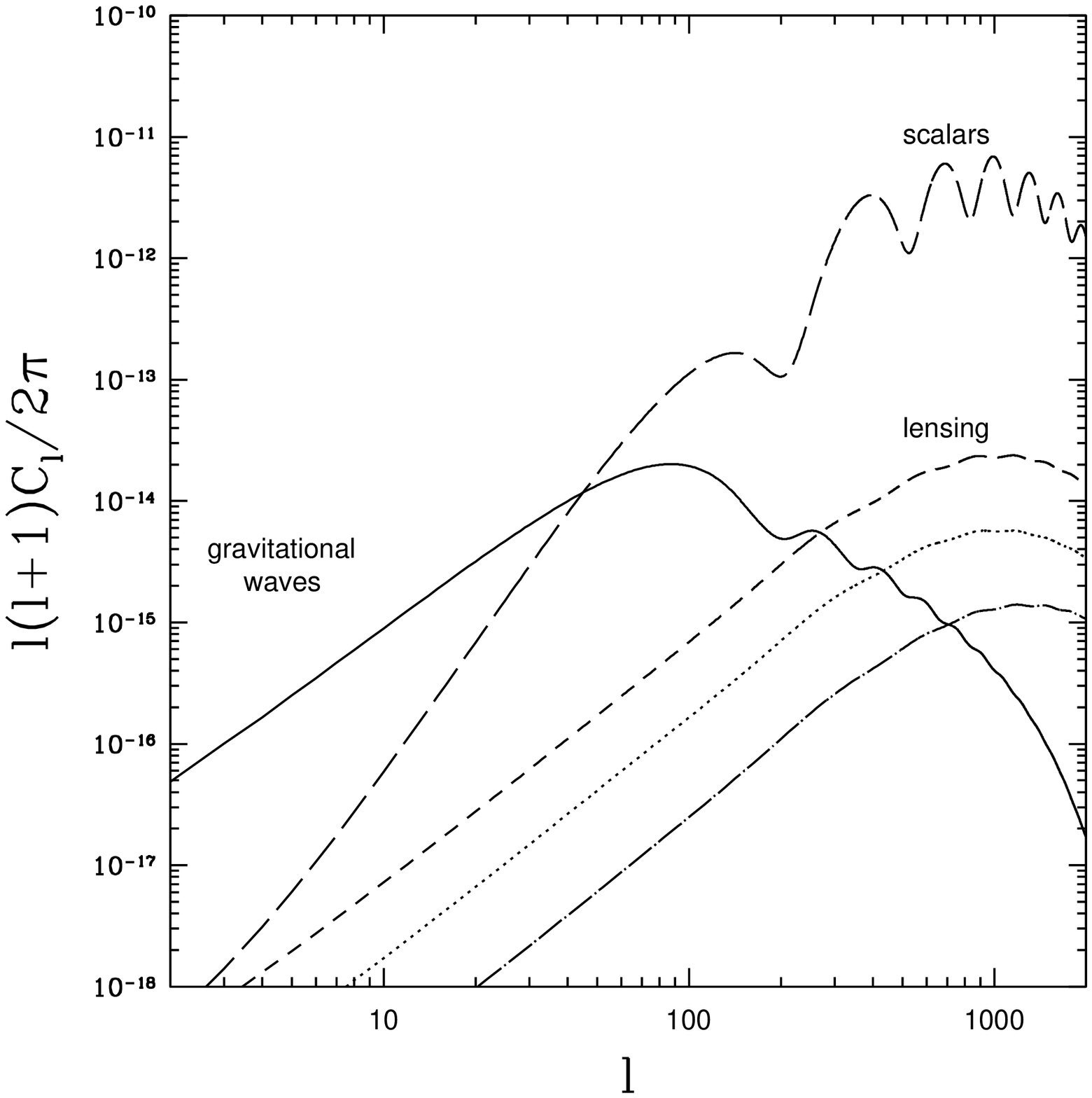}\includegraphics[height=20pc]{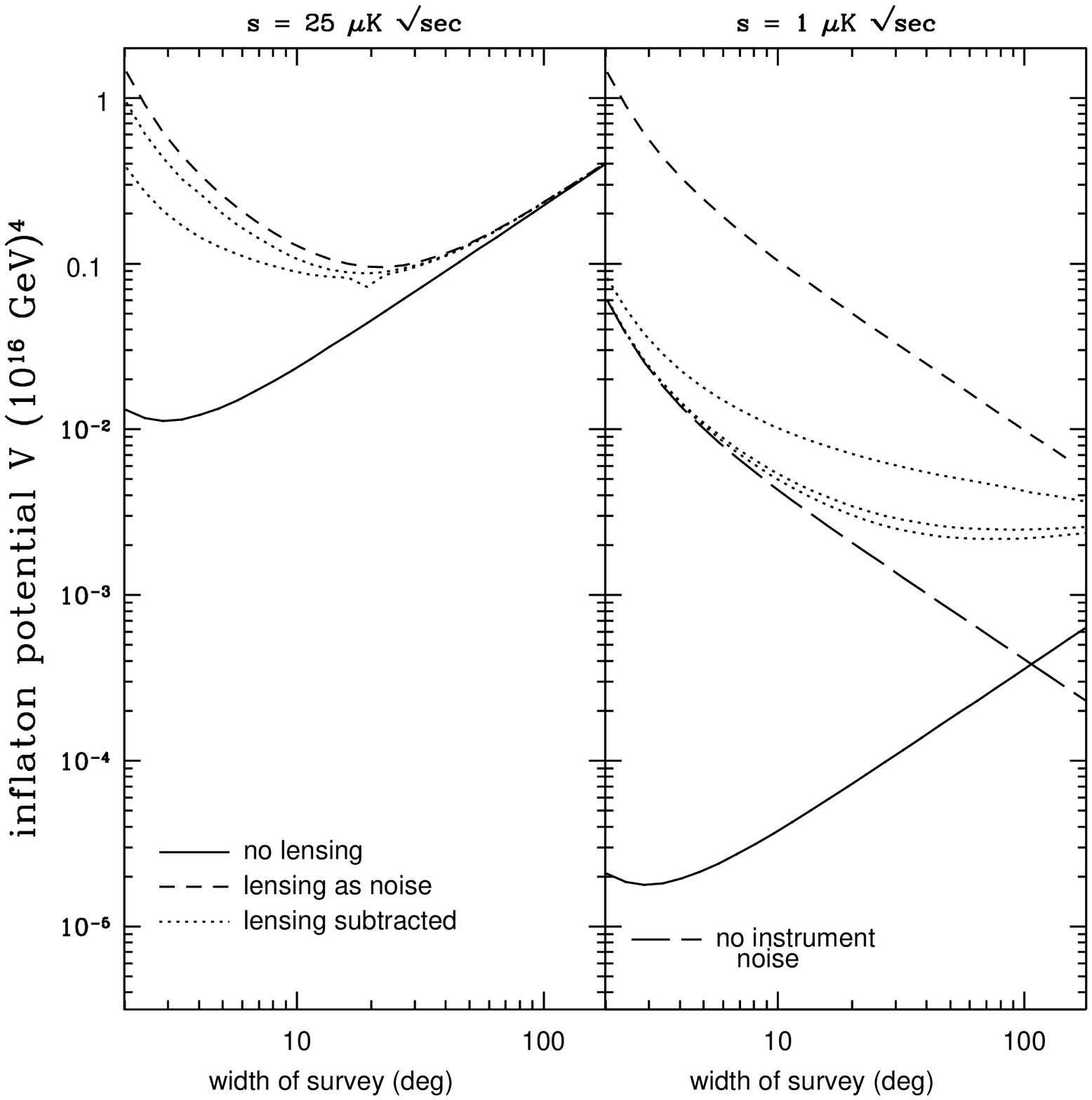}
\caption{{\it Left:}  Contributions to the CMB polarization power spectra.  The
     long-dashed curve shows the dominant polarization signal in the
     gradient component due to scalar perturbations. The solid
     line shows the curl polarization signal from the
     gravitational-wave background assuming inflationary energy scale of
	$2.3 \times 10^{16}$ GeV.
        The dashed curve shows the power
     spectrum of the curl component of the polarization due to
     lensing.  The dotted curve is the lensing contribution to the curl  component that comes from structures out to a redshift of
     1. The dot-dashed line is the residual when
     lensing contribution is separated with a no-noise
     experiment and almost full sky coverage. {\it Right:} Minimum inflation potential observable at
     $1\sigma$ as a function of survey width. The left panel shows an experiment
     with $=25\, \mu{\rm      K}~\sqrt{\rm sec}$ sensitivity.
     The solid curve shows results assuming no lensing
     while the dashed curve shows results including the effects
     of an unsubtracted lensing; we take $\theta_{\rm     FWHM}=5'$ in these two cases.  The dotted curves
     assume lensing is subtracted with $\theta_{\rm     FWHM}=10'$ (upper curve) and $5'$ (lower curve).  The
     right panel shows results 
     assuming $1\, \mu{\rm K}~\sqrt{\rm sec}$ sensitivity and $\theta_{\rm FWHM}=5'$, $2'$, and $1'$     (from top to bottom).
     The solid curve assumes $\theta_{\rm     FWHM}=1'$ and $s=1\, \mu{\rm K}~\sqrt{\rm sec}$, and
     no lensing, while the dashed curve treats lensing as an additional  noise.  The long-dash curve assumes  lensing subtraction
     with no instrumental noise.}
\label{fig:bmodes}
\end{figure}

Though acoustic oscillations in the temperature
anisotropies of CMB suggest an inflationary origin for
primordial perturbations, it has been argued for a while
 that the smoking-gun signature for inflation would be the detection of a stochastic
background of gravitational waves (e.g., Kamionkowski \& Kosowsky 1999).  These gravitational-waves produce a distinct signature in the
CMB in the form of a contribution to the curl, or magnetic-like, component of the polarization (Kamionkowski et al. 1997; Seljak \& Zaldarriaga 1997).
Note that there is no contribution from the dominant scalar, or density-perturbation, contribution to these curl modes.
In figure~\ref{fig:bmodes}, we show the contribution from dominant scalar modes to the polarization in the gradient
 component and from gravitational-waves to the curl polarization. 
There is, however, another source of contribution to the curl component resulting from the 
fractional conversion of the gradient polarization 
via the weak gravitational lensing effect (Zaldarriaga \& Seljak 1998). 
The lensing-induced curl introduces a noise from which gravitational waves must be distinguished in the CMB polarization.
In figure~\ref{fig:bmodes}, we summarize these contributions following Kesden et al. (2002; see, also, Knox \& Song 2002).

As we discussed earlier, when considering the effect of lensing on temperature anisotropies, 
one can use the CMB data, both temperature and polarization, to extract lensing information. This is desirable since replacing the
lensing effect on CMB with what one extract from galaxy shape measurements only leads to a partial accounting of the total effect
(the difference in curves in figure~\ref{fig:convergence}).  By mapping lensing deflection from CMB data with higher order 
correlations, the lensing effect on polarization can be corrected to reconstruct the intrinsic CMB
polarization at the surface of last scatter. Now the only curl component would be that due to gravity waves.
In figure~\ref{fig:bmodes}, we show how well the lensing signal can be extracted. With no noise all-sky map, one can remove the lensing
contribution with a noise contribution that is roughly an order of magnitude lower than the original confusion.
As shown, if there is no instrumental-noise limitation, the sensitivity to gravity-wave
signal is maximized by covering as much sky as possible and allow the accessibility to an inflaton potential of $\sim10^{15}$ GeV.

{\it Acknowledgments:}
The author thanks Wayne Hu and Marc Kamionkowski  for collaborative work and helpful advice, and acknowledges support 
from the Sherman Fairchild foundation and the Department of Energy.


\end{document}